\documentclass[conference]{IEEEtran}
\IEEEoverridecommandlockouts

\usepackage{cite}
\usepackage{amsmath,amssymb,amsfonts}
\usepackage{algorithmic}
\usepackage{graphicx}
\usepackage{textcomp}
\usepackage{xcolor}
\usepackage{booktabs}
\usepackage{multirow}
\usepackage{tipa}
\usepackage{url}
\usepackage{array}
\usepackage{CJKutf8}
\usepackage{eso-pic}

\AddToShipoutPictureBG*{%
    \AtPageLowerLeft{%
        \put(\LenToUnit{0.5\paperwidth},\LenToUnit{0.35in}){%
            \makebox[0pt][c]{\parbox{\textwidth}{\scriptsize\centering
                    \copyright~2026 IEEE. Personal use of this material is permitted. Permission from IEEE must be obtained for all other uses, in any current or future media, including reprinting/republishing this material for advertising or promotional purposes, creating new collective works, for resale or redistribution to servers or lists, or reuse of any copyrighted component of this work in other works.}}}}}

\def\BibTeX{{\rm B\kern-.05em{\sc i\kern-.025em b}\kern-.08em
    T\kern-.1667em\lower.7ex\hbox{E}\kern-.125emX}}

\begin{document}

\title{Phoneme-Guided Initialization for LLM-based Speech Recognition}

\author{\IEEEauthorblockN{Ryo Magoshi, Shinsuke Sakai, and Tatsuya Kawahara}
    \IEEEauthorblockA{Graduate School of Informatics, Kyoto University, Japan \\
        \{magoshi, sakai\}@sap.ist.i.kyoto-u.ac.jp, kawahara@i.kyoto-u.ac.jp}
}

\maketitle

\begin{abstract}
    Speech large language models (speech LLMs) perform well on automatic speech recognition (ASR) when sufficient paired speech-text data is available, but their performance degrades in low-resource settings.
    A cascaded pipeline that performs speech-to-phoneme (S2P) conversion followed by phoneme-to-grapheme (P2G) conversion has been shown to outperform end-to-end speech LLMs in this regime, suggesting that phoneme-mediated processing is beneficial when paired data is scarce.
    We propose \textit{phoneme-guided initialization}, a simple method that uses this insight within an end-to-end framework: we pre-train the audio encoder on S2P and the LLM on P2G tasks, then connect them and fine-tune the full model end-to-end on the target ASR task.
    Experiments on Japanese (CSJ), Chinese (AISHELL-1), and two low-resource languages from Common Voice 25.0 (Tatar and Urdu) show that our method matches or outperforms both the cascaded S2P-P2G baseline and the end-to-end model without P2G initialization.
\end{abstract}

\begin{IEEEkeywords}
    Speech Recognition, Low-Resource, Large Language Model, Phoneme, Initialization
\end{IEEEkeywords}

\section{Introduction}

Augmenting large language models (LLMs) with audio input capabilities has enabled speech LLMs to adapt to various spoken language tasks, including automatic speech recognition (ASR)~\cite{hono-etal-2024-integrating}, speech translation~\cite{prompting2024}, speech-queried question answering~\cite{zhang-etal-2023-speechgpt,audiopalm}, speech synthesis~\cite{valle}, and spoken dialogue~\cite{zhang-etal-2023-speechgpt,audiopalm,dgslm,moshi}.

There are two main approaches to feeding audio into an LLM: discrete and continuous approaches.
The discrete approach quantizes audio into discrete tokens and augments the LLM's text vocabulary with these tokens~\cite{zhang-etal-2023-speechgpt,audiopalm}.
The continuous approach passes the output of a pretrained audio encoder through a lightweight learnable projector to produce continuous representations that are directly processed by the LLM embedding layer~\cite{prompting2024,tang2024salmonn,hu-etal-2024-wavllm}.
This work focuses on the continuous approach, now widely adopted for LLM-based ASR.

Unlike the discrete approach, the continuous approach does not add audio tokens to the vocabulary, but it still introduces a modality mismatch problem~\cite{phonemevsproj}.
In the LLM’s embedding space, closeness of representations corresponds to semantic similarity, whereas in the audio encoder’s output space, closeness of representations corresponds to acoustic similarity.
Although training only the projector can suffice in certain settings~\cite{slam-asr}, end-to-end training of both the projector and the LLM is usually necessary to resolve the modality alignment problem~\cite{slam-ugly}, and even with parameter-efficient techniques such as LoRA~\cite{lora}, a large amount of paired speech-transcription data is still needed for effective cross-modal alignment.
This is a significant limitation, as paired speech-text data is scarce in many conditions, whereas text-only data is more abundant and easier to obtain.

In this work, we propose \textit{phoneme-guided initialization}, an approach that provides LLM-based ASR with initial parameters suited for modality alignment.
Our method uses abundant text data by separately initializing the audio encoder and the LLM through supervised training on phoneme-related tasks before combining them for end-to-end fine-tuning:
\begin{itemize}
    \item The audio encoder is trained on a \textbf{speech-to-phoneme (S2P)} task. Since phonemes are the minimal distinctive sound units of a language, this is simpler than directly predicting graphemes and requires relatively little paired data.
    \item The LLM is trained on a \textbf{phoneme-to-grapheme (P2G)} task, with phoneme sequences generated from external text by grapheme-to-phoneme (G2P) tools, enabling the use of large text corpora without any speech data.
\end{itemize}
After initialization, the encoder and LLM are connected by a projector and fine-tuned end-to-end on the target task.

Our experiments demonstrate that the proposed method outperforms conventional approaches, particularly in low-resource settings in homophone-rich languages.
The key contribution is to show that text-only data from the target language or domain can improve LLM-based ASR through P2G initialization of the LLM, complementing the S2P initialization of the audio encoder so that both components are aligned toward phoneme-level representations before end-to-end training.

\begin{figure}[t]
    \centering
    \includegraphics[width=0.8\columnwidth]{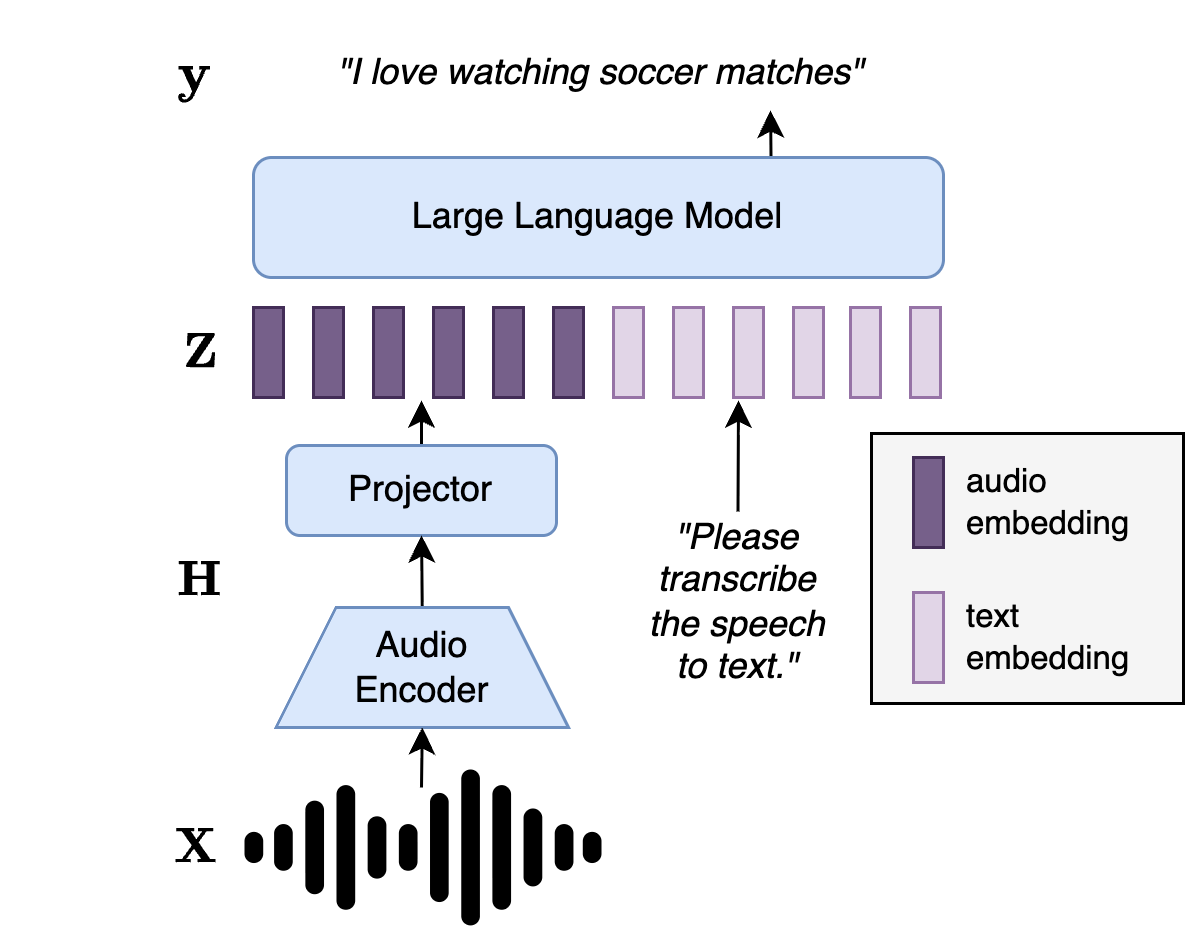}
    \caption{Overview of continuous LLM-based ASR. The audio encoder extracts speech representations, which are mapped to the LLM's embedding space by a projector. The LLM generates the transcription conditioned on both the projected audio and a text prompt. $\mathbf{H}$ is the encoder output and $\mathbf{Z}$ is the projector output (the audio embeddings fed to the LLM).}
    \label{fig:llm_asr}
\end{figure}

\begin{figure*}[t]
    \centering
    \includegraphics[width=0.9\textwidth]{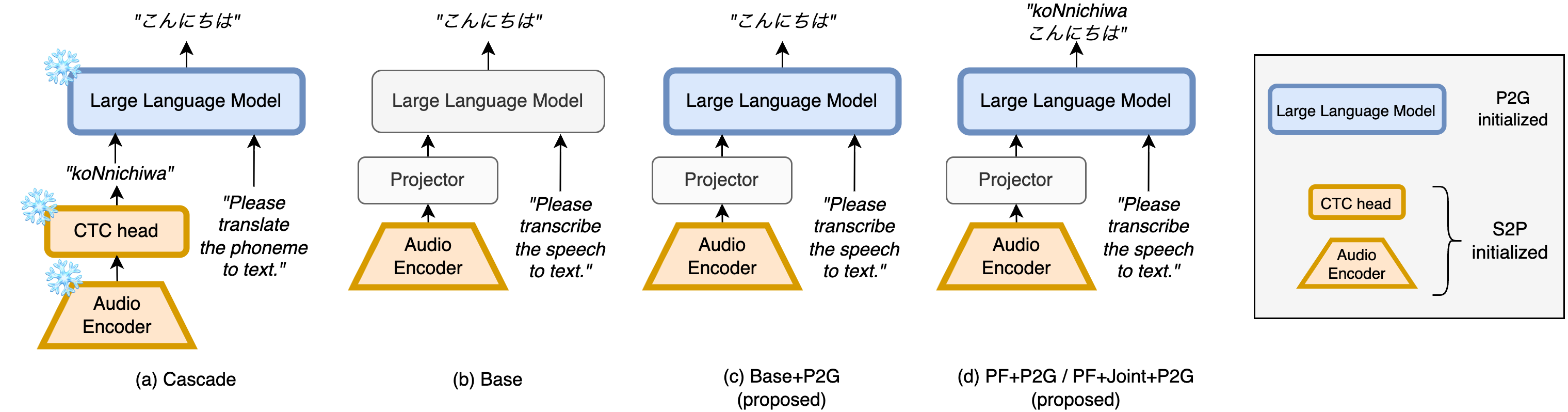}
    \caption{Overview of the LLM-based model configurations, illustrated with the Japanese transcription of ``\begin{CJK}{UTF8}{goth}こんにちは\end{CJK} (koNnichiwa)''. In (a), (c), and (d), the LLM is initialized with P2G training. In (a), all components are frozen and the system operates in a cascaded manner.}
    \label{fig:methods}
\end{figure*}

\section{Related Work}

\subsection{Phonemes in Speech Recognition}

Phonemes are the minimal units of sound that native speakers of a language can distinguish, and phonemic transcription is the highest level of abstraction that still preserves meaning distinctions in a given language.
By discarding paralinguistic and non-linguistic information and mapping sounds that do not carry distinctive meaning to a single symbol, phoneme representations lie closer to linguistic information than the raw speech signal.

The number of distinct phonemes in a language is typically much smaller than the number of graphemes, especially for languages with large character inventories such as Chinese and Japanese, where predicting phonemes from speech is easier than predicting graphemes directly.
Consequently, using phonemic information to predict graphemes has been explored, including in end-to-end systems~\cite{chen19l_interspeech,hu2019phoneme,procter,phoneme_aware_encoding,zhengjie2025pinyin,poncelet2026phoneme,phonemevsproj,yang2024bridging}.

\subsection{Phoneme-Mediated LLM-based ASR}

Jie and Cheng~\cite{zhengjie2025pinyin} investigated phoneme-mediated approaches for Chinese LLM-based ASR using pinyin,
which is a romanization system that serves as a phonemic representation for Mandarin Chinese.
They proposed two methods:
\begin{itemize}
    \item \textbf{Sequential output}: The LLM first generates pinyin, then generates the corresponding Chinese characters.
    \item \textbf{Iterative output}: Pinyin and Chinese characters are interleaved in the output sequence.
\end{itemize}
Both improved over the grapheme-only baseline by using the easier phoneme prediction to guide grapheme prediction.
However, both require the model to output both phoneme and grapheme sequences during training.
Resolving the modality matching problem may then be difficult when data aligning speech, phonemes, and graphemes is scarce.
The iterative method also applies only to languages such as Chinese with a one-to-one grapheme-phoneme correspondence.

Poncelet and Van hamme~\cite{poncelet2026phoneme} adopted a similar sequential output approach and showed improvements on English and Dutch, using ARPAbet for English and Dutch phoneme annotations that were partially manually created.
They further showed that mixing phoneme-first prediction data with standard speech-to-grapheme (S2G) training data, a \textit{joint training} strategy, further improves performance, particularly in low-resource settings with high-quality phoneme annotations.
Since the jointly trained model is trained on both output formats, it can be decoded in either grapheme-only or phoneme-first mode at inference time, and both modes showed improved performance over the non-joint baseline.

Li et al.~\cite{phonemevsproj} investigated LLM-based ASR in low-resource settings.
They found that a cascaded approach, using CTC-based S2P followed by feeding the predicted phonemes into an LLM for P2G, outperforms end-to-end LLM-based ASR.
The CTC-based S2P model was trained in a data-efficient, multilingual manner~\cite{whistle}, which is effective for low-resource settings.
They attributed the poor end-to-end performance to the difficulty of modality matching through the projector in low-resource settings.
They also found that initializing the audio encoder with S2P training reduces this modality mismatch.
However, they did not explore the effect of additionally training the LLM on P2G, which we found effective for further modality alignment.

\subsection{Differences from Prior Work}

Unlike these prior works, which modify the output format of the LLM at inference time, we use phoneme-related tasks exclusively for \textit{initialization}: we pre-train the audio encoder on S2P and the LLM on P2G, then connect them with a projector and fine-tune end-to-end on the target ASR task.
Because only the initialization is modified, the model can be trained with either grapheme-only or phoneme-first output, and we evaluate both configurations.

\section{Methods}

\subsection{LLM-based ASR}

Fig.~\ref{fig:llm_asr} shows the overview of LLM-based ASR.
Given an input speech waveform $\mathbf{X} \in \mathbb{R}^{T_{\text{wav}}}$, an LLM-based ASR system consists of three components: an audio encoder $\text{Enc}(\cdot)$, a projector $\text{Proj}(\cdot)$, and an LLM $\text{LLM}(\cdot)$.

The audio encoder first converts the waveform into a sequence of frame-level representations:
\begin{equation}
    \mathbf{H} = \text{Enc}(\mathbf{X}) \in \mathbb{R}^{T \times d_{\text{enc}}}
\end{equation}
where $T$ is the number of output frames and $d_{\text{enc}}$ is the encoder output dimension.

The projector maps these representations into the LLM's embedding space while downsampling the temporal resolution by a factor of $r$:
\begin{equation}
    \mathbf{Z} = \text{Proj}(\mathbf{H}) \in \mathbb{R}^{\lfloor T/r \rfloor \times d_{\text{llm}}}
\end{equation}
where $d_{\text{llm}}$ is the LLM's hidden dimension.

The projected audio representations $\mathbf{Z}$ are then prepended to the text token embeddings and processed by the LLM. Let $\mathbf{y} = (y_1, y_2, \ldots, y_N)$ denote the target grapheme token sequence. The model is trained to minimize the negative log-likelihood as follows:
\begin{equation}
    \mathcal{L}_{\text{S2G}} = -\sum_{i=1}^{N} \log p(y_i \mid y_{<i}, \mathbf{Z}; \theta)
\end{equation}
where $\theta$ denotes the parameters of the whole model.
At inference time, the output sequence is obtained by beam search, which approximately maximizes the sequence log-likelihood:
\begin{equation}
    \hat{\mathbf{y}} = \arg\max_{\mathbf{y}} \, \sum_{i=1}^{N} \log p(y_i \mid y_{<i}, \mathbf{Z}; \theta)
\end{equation}
The beam size and other decoding settings are given in the training configuration (Section~\ref{sec:experiments}).

\begin{table*}[t]
    \caption{Data configuration for each language. CV 25.0 denotes Common Voice 25.0. ``\# of Sentences'' denotes the number of phoneme-grapheme sentence pairs used for P2G training of the LLM. \\
        For Japanese, the text corpus already contains phoneme transcriptions, so no external G2P tool is needed.}
    \label{tab:data}
    \centering
    \begin{tabular}{lcccccc}
        \toprule
                                  & \multicolumn{2}{c}{\textbf{Speech Data}}        & \multicolumn{2}{c}{\textbf{Text Data (for P2G)}} &                                                                                                                            \\
        \cmidrule(lr){2-3} \cmidrule(lr){4-5}
        \textbf{Language}         & \textbf{Corpus}                                 & \textbf{\# of Hours}                             & \textbf{Corpus}                                   & \textbf{\# of Sentences} & \textbf{G2P Tool}                           \\
        \midrule
        \multirow{2}{*}{Japanese} & \multirow{2}{*}{CSJ~\cite{maekawa2003corpus}}   & 10                                               & \multirow{2}{*}{BCCWJ~\cite{maekawa2014balanced}} & \multirow{2}{*}{1.0M}    & \multirow{2}{*}{--}                        \\
                                  &                                                 & 36                                               &                                                   &                          &                                             \\
        \midrule
        \multirow{3}{*}{Chinese}  & \multirow{3}{*}{AISHELL-1~\cite{bu2017aishell}} & 10                                               & \multirow{3}{*}{THUCNews~\cite{guo2016thuctc}}    & \multirow{3}{*}{2.0M}    & \multirow{3}{*}{g2pW~\cite{chen2022g2pw}}   \\
                                  &                                                 & 50                                               &                                                   &                          &                                             \\
                                  &                                                 & full                                             &                                                   &                          &                                             \\
        \midrule
        Tatar                     & CV 25.0~\cite{ardila2020common}                 & 8.3                                              & Tatar Folklore~\cite{taruen_tatar_folklore}       & 65k                      & Phonetisaurus~\cite{novak2016phonetisaurus} \\
        \midrule
        Urdu                      & CV 25.0~\cite{ardila2020common}                 & 8.4                                              & Makhzan~\cite{makhzan}                            & 211k                     & Epitran~\cite{mortensen2018epitran}         \\
        \bottomrule
    \end{tabular}
\end{table*}

\subsection{Phoneme-First Prediction}

In phoneme-first prediction~\cite{poncelet2026phoneme}, the LLM is trained to first generate a phoneme sequence $\mathbf{p} = (p_1, \ldots, p_M)$ and then the grapheme sequence $\mathbf{y}$, both conditioned on the audio representation. The training objective is:
\begin{equation}
    \mathcal{L}_{\text{PF}} = -\sum_{j=1}^{M} \log p(p_j \mid p_{<j}, \mathbf{Z}; \theta) - \sum_{i=1}^{N} \log p(y_i \mid y_{<i}, \mathbf{p}, \mathbf{Z}; \theta)
\end{equation}
At inference time, the grapheme output is conditioned on both the audio representation and the predicted phonemes, and is likewise decoded by beam search as in Eq.~(4):
\begin{equation}
    \hat{\mathbf{y}} = \arg\max_{\mathbf{y}} \, \sum_{i=1}^{N} \log p(y_i \mid y_{<i}, \hat{\mathbf{p}}, \mathbf{Z}; \theta)
\end{equation}

In the joint training variant~\cite{poncelet2026phoneme}, both phoneme-first and standard S2G training examples are mixed, with a text prompt indicating which output format is expected. This allows the model to learn from both training signals simultaneously.

\subsection{Proposed Method: Phoneme-Guided Initialization}
\label{sec:proposed}

Our proposed method initializes both the audio encoder and the LLM using phoneme-related tasks before end-to-end fine-tuning. The procedure consists of three stages:

\textbf{Stage 1: Component Pre-training.}
\begin{enumerate}
    \item \textbf{S2P initialization of the audio encoder}: The audio encoder is trained with a CTC~\cite{graves2006connectionist} loss to predict phoneme sequences from speech:
          \begin{equation}
              \mathcal{L}_{\text{S2P}} = -\log p_{\text{CTC}}(\mathbf{p} \mid \text{Enc}(\mathbf{X}))
          \end{equation}
          This requires paired speech-phoneme data, but phoneme sequences can be automatically obtained from grapheme transcriptions using G2P tools, and the S2P task is simpler than S2G.

    \item \textbf{P2G initialization of the LLM}: The LLM is fine-tuned on a phoneme-to-grapheme conversion task:
          \begin{equation}
              \mathcal{L}_{\text{P2G}} = -\sum_{i=1}^{N} \log p(y_i \mid y_{<i}, \mathbf{p}; \theta_{\text{LLM}})
          \end{equation}
          where $\mathbf{p}$ is the phoneme sequence generated from the grapheme sequence $\mathbf{y}$ by a G2P tool, and $\theta_{\text{LLM}} \subset \theta$ denotes the LLM parameters. This stage uses only text data and does not require any speech data.
\end{enumerate}

\textbf{Stage 2: Encoder and Projector Warmup.}
The S2P-initialized encoder and P2G-initialized LLM are connected via a randomly initialized projector and fine-tuned on the target ASR task.
The training objective can be either the standard S2G loss $\mathcal{L}_{\text{S2G}}$ (Eq.~3) or the phoneme-first loss $\mathcal{L}_{\text{PF}}$ (Eq.~5); the proposed method concerns only the initialization and imposes no constraint on whether the model outputs graphemes only or phonemes followed by graphemes.
In this stage, only the encoder and projector parameters are updated.

\textbf{Stage 3: Full Fine-tuning.}
All parameters including the LLM are jointly fine-tuned on the same task as Stage~2.
This initialization strategy aligns both the encoder and the LLM toward phoneme-level representations before end-to-end training.

\section{Experiments}
\label{sec:experiments}

\subsection{Model Configuration}

As described in Section~\ref{sec:proposed}, our LLM-based ASR system consists of an audio encoder, a projector, and an LLM.

As the audio encoder, we use XLS-R~\cite{babu2022xls} 300M, a multilingual self-supervised speech representation model.
The projector consists of a temporal downsampling layer with a factor of 4 followed by a two-layer MLP with GeLU~\cite{gelu} activation.
For the LLM, we use Qwen3.5-2B~\cite{qwen3.5}, which originally supports both image and text inputs.
Since we do not use image input in this work, we remove the image encoder and its associated projector from the original model.

We evaluate the following model configurations (Fig.~\ref{fig:methods}):
\begin{itemize}
    \item \textbf{XLS-R + CTC (not in Fig.~\ref{fig:methods})}: The XLS-R encoder is trained with CTC for direct S2G recognition, without an LLM, serving as a non-LLM baseline to isolate the LLM's contribution.
    \item \textbf{Cascade}: The audio encoder is trained with CTC for S2P, and a separately trained LLM performs P2G at inference time; the CTC-predicted phonemes are fed as text input to the LLM. This follows the cascaded approach of~\cite{phonemevsproj}.
    \item \textbf{End-to-end LLM-based ASR}: Following~\cite{phonemevsproj}, all end-to-end models use an audio encoder initialized with S2P training, with the CTC head discarded. We found that without S2P initialization, end-to-end models failed to produce meaningful output (CER near 100\%). There are four subdivided settings:
          \begin{itemize}
              \item \textbf{Base}: The standard end-to-end model trained with the S2G objective. The audio encoder, projector, and LLM are trained following Stages~2 and~3 of Section~\ref{sec:proposed}, with only the S2P part of Stage~1. In this setting, P2G initialization is not conducted.
              \item \textbf{PF} (Phoneme First): The model is trained to first output phonemes and then graphemes, following~\cite{zhengjie2025pinyin,poncelet2026phoneme}. At inference time, the output consists of a phoneme sequence followed by a grapheme sequence, and the error rate is computed only on the grapheme part.
              \item \textbf{PF+Joint}: Phoneme-first training data is mixed with standard S2G data within each batch, following the joint training strategy of~\cite{poncelet2026phoneme}. This enables the model to be decoded with either grapheme-only or phoneme-first prompts at inference time.
              \item \textbf{+P2G (proposed)}: The LLM is initialized with P2G training before end-to-end fine-tuning, as described in Stage~1 of Section~\ref{sec:proposed}. This is our proposed method. The suffix yields \textbf{Base+P2G}, \textbf{PF+P2G}, and \textbf{PF+Joint+P2G}.
          \end{itemize}
\end{itemize}

\subsection{Data Configuration}

Table~\ref{tab:data} summarizes the data used in the experiments.
We conduct experiments on four languages along two scenarios; each language requires a speech corpus for ASR training and a text corpus for P2G initialization of the LLM.
The first scenario is the low-resource \textit{setting}: for Japanese (CSJ~\cite{maekawa2003corpus}) and Chinese (AISHELL-1~\cite{bu2017aishell}), large speech corpora exist but we deliberately limit the training data to small subsets (a 10-hour subset and the 36-hour ``core'' subset for CSJ, and 10-hour, 50-hour, and full subsets for AISHELL-1) to simulate target-domain data scarcity.
The second scenario is the low-resource \textit{language}: for Tatar and Urdu, paired speech-text data is inherently scarce, and we use all available data in Common Voice 25.0~\cite{ardila2020common} (8.3 and 8.4 hours, respectively).

\textbf{Text corpora for P2G.}
The text corpus and its size for each language are listed in Table~\ref{tab:data}: BCCWJ~\cite{maekawa2014balanced} for Japanese, THUCNews~\cite{guo2016thuctc} for Chinese, the Tatar Folklore Text Corpus~\cite{taruen_tatar_folklore} for Tatar, and Makhzan~\cite{makhzan} for Urdu.
We use the full text corpus of each language for P2G training.

\textbf{Phoneme generation.}
To construct phoneme-grapheme pairs for P2G training, phoneme sequences are generated from the text corpora using the per-language G2P tools listed in Table~\ref{tab:data}; BCCWJ already includes phoneme transcriptions, so no external G2P tool is needed for Japanese.
We use a different G2P tool per language according to availability and quality, selecting the best-performing available tool for each. Language-specific phonological rules, including allophonic variation, are handled by the respective G2P tool (or by manual annotation where available, as in BCCWJ), rather than being modeled explicitly.

\subsection{Training Configuration}

Training consists of multiple phases depending on the model configuration.

\textbf{XLS-R + CTC.}
For both S2P and S2G CTC models, we train the XLS-R encoder with a learning rate of $3 \times 10^{-4}$, a batch size of 48, and 1000 warmup steps.
The learning rate is linearly warmed up and then decayed following a cosine schedule.
Throughout all training stages, the CNN feature extractor of the audio encoder is kept frozen.

\textbf{P2G training.}
The LLM is fine-tuned on the P2G task with a learning rate of $5 \times 10^{-5}$, a batch size of 64, and 500 warmup steps.
During P2G training, the input phoneme sequence is provided as a text prompt and the LLM is trained to generate the corresponding grapheme sequence autoregressively.
For the Cascade model, the P2G-trained LLM is further fine-tuned on phoneme-grapheme pairs from the target speech corpus's transcriptions, adapting it to the target domain, which consistently improved P2G performance.
These phoneme sequences come from manually annotated readings for CSJ and from the same G2P tools (Table~\ref{tab:data}) for the other corpora.

\textbf{End-to-end LLM-based ASR.}
End-to-end training follows Stages~2 and~3 of Section~\ref{sec:proposed}.
Stage~2 (encoder and projector only) uses a learning rate of $2 \times 10^{-4}$, a batch size of 24, and 1000 warmup steps; Stage~3 (all parameters) uses $2 \times 10^{-5}$, a batch size of 24, and 500 warmup steps.
For models with the PF (Phoneme First) configuration, the target output sequence is formatted as a concatenation of the phoneme sequence followed by the grapheme sequence.
For PF+Joint, training samples with and without the phoneme prefix are mixed within each batch.

At inference time, we use beam search with beam size 3, repetition penalty 1.0, and prohibited repeated 3-grams.

\subsection{Results}

\begin{table*}[t]
    \caption{Results on four languages. CV denotes Common Voice. \\
        \textbf{Bold} and \underline{underlined} values indicate the best and second best results, respectively. \\
        \textbf{Dec.} denotes the output format at inference: G = grapheme only, P+G = phoneme followed by grapheme. \\
        PF+Joint models support both decoding styles.}
    \label{tab:results}
    \centering
    \setlength{\tabcolsep}{3pt}
    \begin{tabular}{ll*{13}{wc{3em}}}
        \toprule
                                             &               & \multicolumn{8}{c}{\textbf{CSJ (CER $\downarrow$)}} & \multicolumn{3}{c}{\textbf{AISHELL-1 (CER $\downarrow$)}} & \multicolumn{2}{c}{\textbf{CV 25.0 (WER $\downarrow$)}} \\
        \cmidrule(lr){3-10} \cmidrule(lr){11-13} \cmidrule(lr){14-15}
                                             &               & \multicolumn{4}{c}{10h}                             & \multicolumn{4}{c}{core (36h)}                            & 10h              & 50h             & full            & Tatar            & Urdu             \\
        \cmidrule(lr){3-6} \cmidrule(lr){7-10} \cmidrule(lr){11-11} \cmidrule(lr){12-12} \cmidrule(lr){13-13} \cmidrule(lr){14-14} \cmidrule(lr){15-15}
        \textbf{Model}                       & \textbf{Dec.} & eval1 & eval2 & eval3 & avg. & eval1 & eval2 & eval3 & avg. & test & test & test & test & test \\
        \midrule
        XLS-R + CTC                          & G             & 15.5 & \underline{12.3} & 13.5 & 13.8 & \underline{12.5} & \underline{9.7} & 9.9 & \underline{10.7} & 19.9 & 10.1 & 6.5 & 56.0 & 30.0 \\
        \midrule
        Cascade                              & P+G           & 20.0 & 18.0 & 16.4 & 18.1 & 18.9 & 13.2 & 12.4 & 14.8 & 11.4 & 8.1 & 7.0 & 55.8 & 28.8 \\
        \midrule
        Base                                 & G             & 21.2 & 20.2 & 21.0 & 20.8 & 14.8 & 12.6 & 13.0 & 13.5 & 39.1 & 11.2 & 6.5 & \underline{41.6} & 29.3 \\
        PF                                   & P+G           & 24.5 & 23.9 & 24.3 & 24.3 & 16.1 & 13.9 & 14.5 & 14.8 & 18.2 & 9.4 & 6.4 & 46.3 & 33.2 \\
        \multirow{2}{*}{PF+Joint}            & G             & 21.8 & 20.7 & 22.9 & 21.8 & 14.8 & 13.3 & 13.6 & 13.9 & 23.7 & 9.0 & 6.2 & \textbf{40.6} & 29.4 \\
                                             & P+G           & 21.9 & 19.9 & 22.2 & 21.3 & 14.9 & 13.2 & 13.4 & 13.8 & 21.2 & 8.9 & 6.0 & 43.6 & 31.3 \\
        \midrule
        Base+P2G (ours)                      & G             & \textbf{13.4} & \textbf{10.7} & \textbf{11.1} & \textbf{11.7} & \textbf{11.4} & \textbf{9.2} & \textbf{8.8} & \textbf{9.8} & \underline{11.1} & \textbf{5.7} & \textbf{4.1} & \underline{41.6} & \textbf{28.4} \\
        PF+P2G (ours)                        & P+G           & 19.5 & 16.7 & 15.3 & 17.2 & 15.4 & 12.6 & 11.0 & 13.0 & 11.2 & 8.2 & 6.9 & 49.3 & 36.8 \\
        \multirow{2}{*}{PF+Joint+P2G (ours)} & G             & \underline{15.3} & 12.4 & 13.2 & \underline{13.6} & \underline{12.5} & 10.9 & 9.9 & 11.1 & 17.7 & 7.5 & \underline{5.0} & 41.8 & \underline{28.5} \\
                                             & P+G           & 16.3 & 12.9 & \underline{12.4} & 13.8 & 13.4 & 10.9 & \underline{9.5} & 11.2 & \textbf{11.0} & \underline{7.3} & 5.9 & 43.6 & 31.7 \\
        \bottomrule
    \end{tabular}
\end{table*}

Table~\ref{tab:results} presents the results across all languages and settings.
We use character error rate (CER) for CSJ and AISHELL-1, as Japanese and Chinese do not have explicit word boundaries, and word error rate (WER) for Common Voice 25.0 (Tatar and Urdu).
For models with phoneme-first decoding, the error rate is computed on the grapheme part only, removing the phoneme prediction.

We examine the effect of P2G initialization, its interaction with phoneme-first prediction, and the contrast between low-resource settings and low-resource languages.

\textbf{Effect of P2G initialization.}
P2G initialization of the LLM consistently reduced the error rate on CSJ and AISHELL-1.
On CSJ, Base+P2G lowered the average CER from 20.8\% to 11.7\% at 10 hours, a 44\% relative reduction, and by 27\% relative on the 36-hour core set.
The gains were larger on AISHELL-1, where the relative reduction reached 72\% at 10 hours (39.1\% to 11.1\%), 49\% at 50 hours (11.2\% to 5.7\%), and 37\% on the full set (6.5\% to 4.1\%).
The relative reduction grows as the amount of paired speech data decreases, indicating that P2G initialization is most beneficial in low-resource conditions while remaining effective at full scale.

Without P2G initialization, the end-to-end Base model was outperformed by the non-LLM XLS-R + CTC baseline on CSJ, despite its 2B-parameter LLM.
P2G initialization reversed this ordering: Base+P2G outperformed XLS-R + CTC on all CSJ evaluation sets, and outperformed the Cascade model on all languages.
The Cascade was weakest on CSJ, where CTC errors in spontaneous speech propagate to the P2G stage.
On AISHELL-1 at 10 hours it was more competitive (11.4\%), as the phoneme-to-character mapping in Chinese is relatively well defined, yet Base+P2G (11.1\%) still attained a lower error rate.

Base+P2G is additionally trained on target-language text during P2G initialization, so phoneme-level alignment and plain text adaptation are not fully isolated here.
We therefore leave a matched text-only control to future work.
Still, P2G shifts the error profile toward sound-faithful confusions, the same direction observed when more paired speech-text data is added (Table~\ref{tab:homophone}, Base rows).
This shift is more consistent with improved acoustic-to-text alignment than with better language modeling.
The LLM is also already competent in Chinese and Japanese, limiting the gains available from added language knowledge.

\textbf{Interaction with phoneme-first prediction.}
The effect of phoneme-first prediction was language-dependent.
Without P2G initialization, PF improved over Base on AISHELL-1, consistent with the Chinese results reported by Jie and Cheng~\cite{zhengjie2025pinyin}, but degraded performance on CSJ.
This is likely because each Chinese character corresponds to exactly one syllable, which strongly constrains character prediction, whereas kanji readings are variable-length and context-dependent.
When combined with P2G initialization, phoneme-first prediction consistently hurt performance (PF+P2G $<$ Base+P2G on all settings), suggesting that P2G initialization already provides sufficient phoneme knowledge and makes explicit phoneme output redundant.
PF+Joint+P2G fell between the two, with grapheme-only decoding competitive with Base+P2G and phoneme-first decoding achieving the best AISHELL-1 10-hour result (11.0\%).

\textbf{Low-resource setting vs.\ low-resource language.}
The two evaluation scenarios revealed different behaviors.
For the low-resource \textit{setting} scenario (CSJ and AISHELL-1), where we limited the speech data for well-resourced languages, P2G initialization was effective.
For the low-resource \textit{language} scenario (Tatar and Urdu), the improvement was limited: Base+P2G showed no improvement on Tatar (41.6\%) and only a modest improvement on Urdu (29.3\% to 28.4\%).
This discrepancy likely has two causes.
First, the G2P tools for these low-resource languages may produce lower-quality phonemes than the transcriptions in BCCWJ or the specialized g2pW tool for Chinese.
Second, the writing systems of Tatar (Cyrillic, a full alphabet marking all vowels and consonants, albeit with some ambiguities inherited from Russian) and Urdu (a segmental abjad that omits short vowels) are more phonemic than logographic Chinese or kanji-kana Japanese, leaving a larger phoneme-grapheme gap for the latter.
For more phonemic orthographies, P2G initialization thus provides less additional information to the LLM.
Notably, for Tatar the best-performing model was PF+Joint without P2G initialization (40.6\%), suggesting that joint training itself can be more beneficial than P2G initialization when text resources are scarce and the orthography is already phonemic.

\subsection{Error Analysis}

We also analyzed the nature of errors by examining homophone substitutions on AISHELL-1 and CSJ.
We compute the \textit{homophone share} of substitution errors: among all scorable substitutions (those in which the reference and hypothesis graphemes both have a defined reading), the fraction in which the two graphemes share a reading, i.e., the model predicted the correct sound but the wrong surface form.
Reference and hypothesis are aligned by minimum edit distance, and readings are compared using toneless pinyin for Mandarin and kana for Japanese; a high share indicates that the residual errors are sound-faithful rather than hallucinated.
Table~\ref{tab:homophone} reports this share for all six end-to-end systems.

Across every objective variant and both homophone-rich languages, P2G initialization raises the homophone share (Table~\ref{tab:homophone}), and all fifteen Mandarin and Japanese comparisons are positive.
This confirms quantitatively that P2G initialization shifts the error profile from semantic-drift errors, in which the Base model produces fluent but phonetically unrelated words, toward sound-faithful confusions in which the correct sound is preserved but the wrong character is chosen; Table~\ref{tab:error_examples} gives representative examples of each type.

The magnitude of this shift is informative: on AISHELL-1 the P2G increment shrinks as paired speech data grows, while the Base share alone increases.
Additional paired supervision and P2G initialization thus shift the error profile in the same direction, and P2G reaches a comparable profile from text alone; this is consistent with P2G substituting for paired data and explains why its benefit is largest in the lowest-resource setting.
The effect size also tracks how much homophony each script carries (Mandarin $\gg$ Japanese $\gg$ Urdu $\approx$ Tatar): for Tatar and Urdu the share is low and essentially unchanged by P2G, consistent with the limited ASR gains on these more phonemic orthographies.
The residual errors after P2G are therefore primarily an orthographic disambiguation problem rather than an acoustic one.
Because tone is not distinguished by toneless pinyin, Mandarin near-homophones that differ only in tone (e.g., \begin{CJK}{UTF8}{gbsn}至\end{CJK}~\textit{zh\`{\i}} vs.\ \begin{CJK}{UTF8}{gbsn}之\end{CJK}~\textit{zh\={\i}}) are counted as homophones in the reported share; accurately recovering lexical tone is therefore a remaining limitation of the current approach.

\begin{table}[t]
    \caption{Homophone (sound-faithful) share of substitution errors (\%). Readings use toneless pinyin for Mandarin (character level), kana for Japanese (word level), and the grapheme reading for Tatar and Urdu (word level). Higher is more sound-faithful; compare each \textbf{+P2G} row with the row above it.}
    \label{tab:homophone}
    \centering
    \setlength{\tabcolsep}{4pt}
    \small
    \begin{tabular}{lccccccc}
        \toprule
                             & \multicolumn{3}{c}{\textbf{AISHELL-1 (zh)}} & \multicolumn{2}{c}{\textbf{CSJ (ja)}} & \textbf{Ur} & \textbf{Tt}                                     \\
        \cmidrule(lr){2-4} \cmidrule(lr){5-6}
        \textbf{System}      & 10h                                         & 50h                                   & full        & 10h         & core & test & test \\
        \midrule
        Base                 & 13.2                                        & 37.4                                  & 49.7        & 5.3         & 7.4  & 2.8  & 0.8  \\
        Base+P2G             & 37.8                                        & 50.1                                  & 54.1        & 9.4         & 10.2 & 2.8  & 0.8  \\
        \midrule
        PF                   & 38.5                                        & 44.9                                  & 48.8        & 4.6         & 7.6  & 2.8  & 0.8  \\
        PF+P2G               & 50.8                                        & 53.3                                  & 53.9        & 6.9         & 10.3 & 2.7  & 0.7  \\
        \midrule
        PF+Joint             & 26.9                                        & 42.8                                  & 50.7        & 5.8         & 7.2  & 2.9  & 0.8  \\
        PF+Joint+P2G         & 43.1                                        & 52.4                                  & 54.1        & 10.8        & 9.7  & 2.8  & 0.7  \\
        \bottomrule
    \end{tabular}
\end{table}

\begin{table}[t]
    \caption{Examples of substitution errors (10 hours). Homophone errors share the reference's reading; semantic-drift errors differ in sound. Mandarin near-homophones match segmentally but may differ in tone.}
    \label{tab:error_examples}
    \centering
    \small
    \begin{tabular}{llll}
        \toprule
        \textbf{Model} & \textbf{Reference}                                          & \textbf{Prediction}                                         & \textbf{Type}  \\
        \midrule
        \multicolumn{4}{l}{\textit{AISHELL-1}}                                                                                                                      \\
        \midrule
        Base           & \begin{CJK}{UTF8}{gbsn}至关\end{CJK} \textit{zh\`{\i}-gu\=an} & \begin{CJK}{UTF8}{gbsn}监管\end{CJK} \textit{ji\=an-gu\v{a}n} & Semantic-drift \\
        Base+P2G       & \begin{CJK}{UTF8}{gbsn}至\end{CJK} \textit{zh\`{\i}}         & \begin{CJK}{UTF8}{gbsn}之\end{CJK} \textit{zh\={\i}}         & Near-homophone \\
        \midrule
        \multicolumn{4}{l}{\textit{CSJ}}                                                                                                                            \\
        \midrule
        Base           & \begin{CJK}{UTF8}{min}近似\end{CJK} \textit{kinji}            & \begin{CJK}{UTF8}{min}認識\end{CJK} \textit{ninshiki}         & Semantic-drift \\
        Base+P2G       & \begin{CJK}{UTF8}{min}回帰\end{CJK} \textit{kaiki}            & \begin{CJK}{UTF8}{min}快気\end{CJK} \textit{kaiki}            & Homophone      \\
        \bottomrule
    \end{tabular}
\end{table}

\section{Conclusion}

We have proposed phoneme-guided initialization for LLM-based ASR, which pre-trains the audio encoder on S2P and the LLM on P2G before end-to-end fine-tuning, pre-aligning the LLM toward phoneme-level representations without any paired speech data.
On Japanese and Chinese it reduced CER by 27 to 72\% relative to the end-to-end baseline and outperformed both the CTC baseline and the cascaded S2P-P2G pipeline, with the largest gains in low-resource settings.
Error analysis showed that P2G shifts the error profile from semantic-drift errors to sound-faithful homophone confusions, indicating that residual errors are primarily orthographic rather than acoustic.
On Tatar and Urdu the gains were limited but not negative, likely because their more phonemic orthographies leave a smaller phoneme-grapheme gap; future work includes higher-quality G2P, larger text corpora, and initialization beyond phoneme-level alignment.

\section*{Acknowledgment}
The authors used generative AI tools only to improve the manuscript's language, and are responsible for its content.

\clearpage
\bibliographystyle{IEEEtran}
\bibliography{refs}

\begin{thebibliography}{10}
\providecommand{\url}[1]{#1}
\csname url@samestyle\endcsname
\providecommand{\newblock}{\relax}
\providecommand{\bibinfo}[2]{#2}
\providecommand{\BIBentrySTDinterwordspacing}{\spaceskip=0pt\relax}
\providecommand{\BIBentryALTinterwordstretchfactor}{4}
\providecommand{\BIBentryALTinterwordspacing}{\spaceskip=\fontdimen2\font plus
\BIBentryALTinterwordstretchfactor\fontdimen3\font minus \fontdimen4\font\relax}
\providecommand{\BIBforeignlanguage}[2]{{%
\expandafter\ifx\csname l@#1\endcsname\relax
\typeout{** WARNING: IEEEtran.bst: No hyphenation pattern has been}%
\typeout{** loaded for the language `#1'. Using the pattern for}%
\typeout{** the default language instead.}%
\else
\language=\csname l@#1\endcsname
\fi
#2}}
\providecommand{\BIBdecl}{\relax}
\BIBdecl

\bibitem{hono-etal-2024-integrating}
Y.~Hono, K.~Mitsuda, T.~Zhao, K.~Mitsui, T.~Wakatsuki, and K.~Sawada, ``Integrating pre-trained speech and language models for end-to-end speech recognition,'' in \emph{Findings of ACL}, 2024, pp. 13\,289--13\,305.

\bibitem{prompting2024}
Y.~Fathullah, C.~Wu, E.~Lakomkin, J.~Jia, Y.~Shangguan, K.~Li, J.~Guo, W.~Xiong, J.~Mahadeokar, O.~Kalinli, C.~Fuegen, and M.~Seltzer, ``Prompting large language models with speech recognition abilities,'' in \emph{Proc. ICASSP}, 2024, pp. 13\,351--13\,355.

\bibitem{zhang-etal-2023-speechgpt}
D.~Zhang, S.~Li, X.~Zhang, J.~Zhan, P.~Wang, Y.~Zhou, and X.~Qiu, ``{S}peech{GPT}: Empowering large language models with intrinsic cross-modal conversational abilities,'' in \emph{Findings of EMNLP}, 2023, pp. 15\,757--15\,773.

\bibitem{audiopalm}
P.~K. Rubenstein, C.~Asawaroengchai, D.~D. Nguyen, A.~Bapna, Z.~Borsos, F.~de~Chaumont~Quitry, P.~Chen, D.~El~Badawy, W.~Han, E.~Kharitonov, H.~Muckenhirn, D.~Padfield, J.~Qin, D.~Rozenberg, T.~Sainath, J.~Schalkwyk, M.~Sharifi, M.~Tadmor~Ramanovich, M.~Tagliasacchi, A.~Tudor, M.~Velimirovi{\'c}, D.~Vincent, J.~Yu, Y.~Wang, V.~Zayats, N.~Zeghidour, Y.~Zhang, Z.~Zhang, L.~Zilka, and C.~Frank, ``{A}udio{PaLM}: A large language model that can speak and listen,'' 2023, arXiv:2306.12925.

\bibitem{valle}
S.~Chen, C.~Wang, Y.~Wu, Z.~Zhang, L.~Zhou, S.~Liu, Z.~Chen, Y.~Liu, H.~Wang, J.~Li, L.~He, S.~Zhao, and F.~Wei, ``Neural codec language models are zero-shot text to speech synthesizers,'' \emph{TASLP}, vol.~33, pp. 705--718, 2025.

\bibitem{dgslm}
T.~A. Nguyen, E.~Kharitonov, J.~Copet, Y.~Adi, W.-N. Hsu, A.~Elkahky, P.~Tomasello, R.~Algayres, B.~Sagot, A.~Mohamed, and E.~Dupoux, ``Generative spoken dialogue language modeling,'' \emph{TACL}, vol.~11, pp. 250--266, 2023.

\bibitem{moshi}
A.~D{\'e}fossez, L.~Mazar{\'e}, M.~Orsini, A.~Royer, P.~P{\'e}rez, H.~J{\'e}gou, E.~Grave, and N.~Zeghidour, ``Moshi: A speech-text foundation model for real-time dialogue,'' 2024, arXiv:2410.00037.

\bibitem{tang2024salmonn}
C.~Tang, W.~Yu, G.~Sun, X.~Chen, T.~Tan, W.~Li, L.~Lu, Z.~Ma, and C.~Zhang, ``{SALMONN}: Towards generic hearing abilities for large language models,'' in \emph{Proc. ICLR}, 2024.

\bibitem{hu-etal-2024-wavllm}
S.~Hu, L.~Zhou, S.~Liu, S.~Chen, L.~Meng, H.~Hao, J.~Pan, X.~Liu, J.~Li, S.~Sivasankaran, L.~Liu, and F.~Wei, ``{W}av{LLM}: Towards robust and adaptive speech large language model,'' in \emph{Findings of EMNLP}, 2024, pp. 4552--4572.

\bibitem{phonemevsproj}
Z.~Li, L.~Dong, S.~Yusuyin, X.~Zhao, and Z.~Ou, ``Phonemes vs. projectors: An investigation of speech-language interfaces for {LLM}-based {ASR},'' 2026, arXiv:2604.09332.

\bibitem{slam-asr}
Z.~Ma, G.~Yang, Y.~Yang, Z.~Gao, J.~Wang, Z.~Du, F.~Yu, Q.~Chen, S.~Zheng, S.~Zhang, and X.~Chen, ``An embarrassingly simple approach for {LLM} with strong {ASR} capacity,'' 2024, arXiv:2402.08846.

\bibitem{slam-ugly}
S.~Kumar, I.~Thorbecke, S.~Burdisso, E.~Villatoro-Tello, M.~K~E, K.~Hacioglu, P.~Rangappa, P.~Motlicek, A.~Ganapathiraju, and A.~Stolcke, ``Performance evaluation of {SLAM-ASR}: The good, the bad, the ugly, and the way forward,'' in \emph{Proc. ICASSP Workshops}, 2025, pp. 1--5.

\bibitem{lora}
E.~J. Hu, Y.~Shen, P.~Wallis, Z.~Allen-Zhu, Y.~Li, S.~Wang, L.~Wang, and W.~Chen, ``{L}o{RA}: Low-rank adaptation of large language models,'' in \emph{Proc. ICLR}, 2022.

\bibitem{chen19l_interspeech}
Z.~Chen, M.~Jain, Y.~Wang, M.~L. Seltzer, and C.~Fuegen, ``{Joint Grapheme and Phoneme Embeddings for Contextual End-to-End ASR},'' in \emph{{Proc. Interspeech}}, 2019, pp. 3490--3494.

\bibitem{hu2019phoneme}
K.~Hu, A.~Bruguier, T.~N. Sainath, R.~Prabhavalkar, and G.~Pundak, ``Phoneme-based contextualization for cross-lingual speech recognition in end-to-end models,'' \emph{arXiv preprint arXiv:1906.09292}, 2019.

\bibitem{procter}
R.~Pandey, R.~Ren, Q.~Luo, J.~Liu, A.~Rastrow, A.~Gandhe, D.~Filimonov, G.~Strimel, A.~Stolcke, and I.~Bulyko, ``Procter: Pronunciation-aware contextual adapter for personalized speech recognition in neural transducers,'' in \emph{Proc. ICASSP}, 2023, pp. 1--5.

\bibitem{phoneme_aware_encoding}
H.~Futami, E.~Tsunoo, Y.~Kashiwagi, H.~Ogawa, S.~Arora, and S.~Watanabe, ``Phoneme-aware encoding for prefix-tree-based contextual asr,'' in \emph{Proc. ICASSP}, 2024, pp. 10\,641--10\,645.

\bibitem{zhengjie2025pinyin}
Z.~Jie and G.~Cheng, ``Pinyin-guided {C}hinese speech recognition with large language model,'' in \emph{Proc. Interspeech}, 2025, pp. 564--568.

\bibitem{poncelet2026phoneme}
J.~Poncelet and H.~Van~hamme, ``Phoneme-first prediction for {LLM}-based speech recognition,'' 2026, arXiv:2606.10864.

\bibitem{yang2024bridging}
Y.~Yang, Y.~Peng, E.~S. Chng, and X.~Zhong, ``Bridging speech and text: Enhancing {ASR} with pinyin-to-character pre-training in {LLM}s,'' in \emph{Proc. 14th International Symposium on Chinese Spoken Language Processing (ISCSLP)}, 2024, pp. 646--650.

\bibitem{whistle}
S.~Yusuyin, T.~Ma, H.~Huang, W.~Zhao, and Z.~Ou, ``Whistle: Data-efficient multilingual and crosslingual speech recognition via weakly phonetic supervision,'' \emph{TASLP}, vol.~33, pp. 1440--1453, 2025.

\bibitem{maekawa2003corpus}
K.~Maekawa, ``Corpus of spontaneous {J}apanese: Its design and evaluation,'' in \emph{Proc. ISCA/IEEE Workshop on Spontaneous Speech Processing and Recognition}, 2003.

\bibitem{maekawa2014balanced}
K.~Maekawa, M.~Yamazaki, T.~Ogiso, T.~Maruyama, H.~Ogura, W.~Kashino, H.~Koiso, M.~Yamaguchi, M.~Tanaka, and Y.~Den, ``Balanced corpus of contemporary written {J}apanese,'' \emph{Language Resources and Evaluation}, vol.~48, no.~2, pp. 345--371, 2014.

\bibitem{bu2017aishell}
H.~Bu, J.~Du, X.~Na, B.~Wu, and H.~Zheng, ``{AISHELL}-1: An open-source {M}andarin speech corpus and a speech recognition baseline,'' in \emph{Proc. O-COCOSDA}, 2017, pp. 1--5.

\bibitem{guo2016thuctc}
Z.~Guo, Y.~Zhao, Y.~Zheng, X.~Si, Z.~Liu, and M.~Sun, ``{THUCTC}: An efficient {C}hinese text classifier,'' \url{http://thuctc.thunlp.org/}, 2016.

\bibitem{chen2022g2pw}
Y.-C. Chen, Y.-C. Steven, Y.-C. Chang, and Y.-R. Yeh, ``g2p{W}: A conditional weighted softmax {BERT} for polyphone disambiguation in {M}andarin,'' in \emph{Proc. Interspeech}, 2022, pp. 1926--1930.

\bibitem{ardila2020common}
R.~Ardila, M.~Branson, K.~Davis, M.~Kohler, J.~Meyer, M.~Henretty, R.~Morais, L.~Saunders, F.~Tyers, and G.~Weber, ``Common voice: A massively-multilingual speech corpus,'' in \emph{Proc. LREC}, 2020, pp. 4218--4222.

\bibitem{taruen_tatar_folklore}
{Taruen}, ``Tatar folklore text corpus,'' Mozilla Data Collective, \url{https://mozilladatacollective.com/datasets/cml8gixh60087o407lfgoumgu}, n.d.

\bibitem{novak2016phonetisaurus}
J.~R. Novak, N.~Minematsu, and K.~Hirose, ``Phonetisaurus: Exploring grapheme-to-phoneme conversion with joint n-gram models in the {WFST} framework,'' \emph{Natural Language Engineering}, vol.~22, no.~6, pp. 907--938, 2016.

\bibitem{makhzan}
Z.~Ahmed \emph{et~al.}, ``Makhzan: An {U}rdu text corpus,'' \url{https://github.com/zeerakahmed/makhzan/}, n.d.

\bibitem{mortensen2018epitran}
D.~R. Mortensen, S.~Dalmia, and P.~Littell, ``{E}pitran: Precision {G2P} for many languages,'' in \emph{Proc. LREC}, 2018.

\bibitem{graves2006connectionist}
A.~Graves, S.~Fern{\'a}ndez, F.~Gomez, and J.~Schmidhuber, ``Connectionist temporal classification: Labelling unsegmented sequence data with recurrent neural networks,'' in \emph{Proc. ICML}, 2006, pp. 369--376.

\bibitem{babu2022xls}
A.~Babu, C.~Wang, A.~Tjandra, K.~Lakhotia, Q.~Xu, N.~Goyal, K.~Singh, P.~von Platen, Y.~Saraf, J.~Pino, A.~Baevski, A.~Conneau, and M.~Auli, ``{XLS-R}: Self-supervised cross-lingual speech representation learning at scale,'' in \emph{Proc. Interspeech}, 2022, pp. 2278--2282.

\bibitem{gelu}
D.~Hendrycks and K.~Gimpel, ``Gaussian error linear units ({GELUs}),'' 2016, arXiv:1606.08415.

\bibitem{qwen3.5}
{Qwen Team}, ``{Qwen3.5}: Towards native multimodal agents,'' \url{https://qwen.ai/blog?id=qwen3.5}, 2026.

\end{thebibliography}

\end{document}